\documentclass{ws-procs9x6}
\newcommand*{\be}{\begin{equation}}
\newcommand*{\ee}{\end{equation}}

\newcommand*{\bea}{\begin{eqnarray}}
\newcommand*{\eea}{\end{eqnarray}}

\newcommand*{\pd}{\partial}
\newcommand*{\pdm}{\pd_{\mu}}
\newcommand*{\pdn}{\pd_{\nu}}

\begin{document}

\title{Residual Confinement\\ in High-Temperature 
Yang-Mills Theory\footnote{\uppercase{T}his 
work is supported by the \uppercase{BMBF} under grant number 06\uppercase{DA}917, 
by the \uppercase{E}uropean \uppercase{G}raduate  \uppercase{S}chool  
\uppercase{B}asel-\uppercase{T}\"ubingen (\uppercase{DFG} contract  \uppercase{GRK}683), 
and by the
\uppercase{H}elmholtz association (\uppercase{V}irtual \uppercase{T}heory 
\uppercase{I}nstitute \uppercase{VH}-\uppercase{VI}-041).}}

\author{A. Maas, J. Wambach}

\address{Darmstadt University of Technology, Schlo{\ss}gartenstra{\ss}e 9, D-64289 
Darmstadt, Germany}

\author{B. Gr\"uter, R. Alkofer}

\address{T\"ubingen University, Auf der Morgenstelle 14, D-72076 T\"ubingen, Germany}  

\maketitle

\abstracts{
The infrared behavior of Landau gauge gluon and ghost propagators are
investigated in Yang-Mills theory at non-vanishing temperatures. 
Self-consistent solutions are presented for temperatures below  the presumed
phase transition and in the infinite temperature limit.  Gluon confinement is
manifest in the infrared behavior of these propagators. As expected confinement
prevails below the phase transition. In the  infinite-temperature limit a
qualitative change is observed: the chromoelectric  sector exhibits a
near-perturbative behavior while long-range chromomagnetic  interactions,
mediated by soft ghost modes,  are still present. The latter behavior is in
agreement with corresponding  lattice results. It furthermore implies that part
of the gluons are still confined.}

\section{Introduction}

It is by now well established that QCD undergoes a phase transition or at least
a  rapid crossover at some critical temperature. The naive expectation is that 
the novel kind of matter formed above the phase transition is a  plasma of
deconfined gluons and quarks reaching ideal gas behavior at  very high
temperatures. Recent lattice calculations (see {\it e.g.\/}
Cucchieri et al.\cite{Cucchieri:2001tw})  have raised doubts about this simple picture. In
addition, it is known that the dimensionally  reduced theory, to which the
equilibrium state evolves in the infinite temperature limit, is confining. The
present work further investigates the properties of finite- and 
high-temperature QCD within the Dyson-Schwinger approach\cite{Alkofer:2000wg}.

The following investigations are restricted to pure Yang-Mills theory as
substantial evidence  exists that the non-perturbative features of QCD are
generated in the gauge sector.  Hence, the theory studied here is an
equilibrium Yang-Mills theory governed by the  Euclidean Lagrangian
\bea
\mathcal{L}&=\frac{1}{4}F_{\mu\nu}^aF_{\mu\nu}^a+\bar c^a \pdm D_\mu^{ab} c^b\nonumber\\
F^a_{\mu\nu}&=\pdm A_\nu^a-\pdn A_\mu^a-gf^{abc}A_\mu^bA_\nu^c\nonumber\\
D_\mu^{ab}&=\delta^{ab}\pdm+gf^{abc}A_\mu^c\nonumber~,
\eea
with the field strength tensors $F_{\mu\nu}^a$ and the covariant derivative 
$D_\mu^{ab}$. $A_\mu^a$ denotes the gluon field and $\bar c^a$ and $c^a$  are 
the ghost fields describing part of the quantum
fluctuations of the gluon  field. The gauge chosen is the Landau gauge which is
best suited for the purpose at hand.

The gluon and ghost fields are described by their respective two-point Green's 
functions or propagators. Their infrared properties are linked to the presence 
of confinement. Especially,  a particle is absent from the 
physical spectrum if its propagator $D(q^2)$ 
vanishes in the infrared\cite{Alkofer:2000wg} 
\be
\lim_{q\to 0} D(q^2)=0.
\label{confcrit}
\ee

\section{Results}

\begin{figure}[ht]
\centerline{\epsfxsize=4.1in\epsfbox{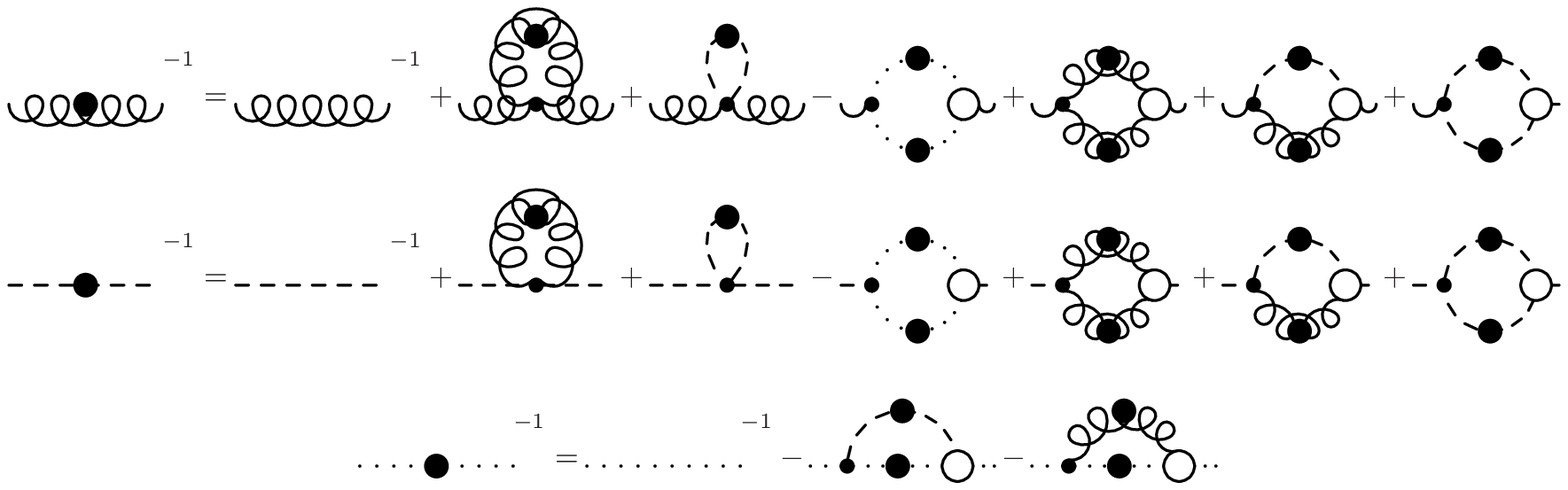}}   
\caption{The truncated Dyson-Schwinger equations at finite temperature. 
The dotted lines denote ghosts, the dashed lines longitudinal gluons and the wiggly 
lines transverse gluons. Lines with a full dot represent self-consistent propagators 
and small dots indicate bare vertices. The open circled vertices are full and must 
be constructed in a given truncation scheme. A bare ghost-gluon vertex and slightly 
modified bare gluon vertices have been used. Note that for soft modes the ghost-longitudinal 
and the 3-point coupling of three longitudinal and of one longitudinal and two 
transverse gluons vanish.}\label{figsysft}
\end{figure}

To present results we first define dressing functions from the propagators:
\bea
&D_G(q)=-\frac{G(q_0^2,q_3^2)}{q^2} \; ,\nonumber\\
&D_{\mu\nu}(q)=P_{\mu\nu}^T(q)\frac{Z(q_0^2,q_3^2)}{q^2}+P_{\mu\nu}^L(q)
\frac{H(q_0^2,q_3^2)}{q^2}\nonumber~.
\eea
Here $G$ denotes the ghost dressing function, $Z$ and $H$ the ones for gluons 
being transverse or longitudinal with respect to the heat bath. To obtain
these  functions the truncated set of Dyson-Schwinger equations  displayed in
Fig.~\ref{figsysft}   have been solved\cite{Fischer:2002hn,torus,d3}.  The
necessity to truncate the infinite set of coupled  Dyson-Schwinger equations
generates several problems with  gauge invariance. These
have been dealt with and the errors are, at least qualitatively,  under
control.

Calculations have been performed at $T=0$\cite{Fischer:2002hn},
$T<T_c$\cite{torus}  and $T\to\infty$\cite{d3}. Those at $T<T_c$ have used a
toroidal  discretized space while the others were done in the continuum. Note
that,  for $T\to\infty$, the theory becomes dimensionally reduced to a 
three-dimensional (3d) Yang-Mills theory  plus an adjoint Higgs field, where
$Z$ corresponds to the 3d gluon (the chromomagnetic  sector of the original
4d theory) and $H$ to the Higgs (the chromoelectric sector  of the original
4d theory).

\begin{figure}[ht]
\centerline{\epsfxsize=2.3in\epsfbox{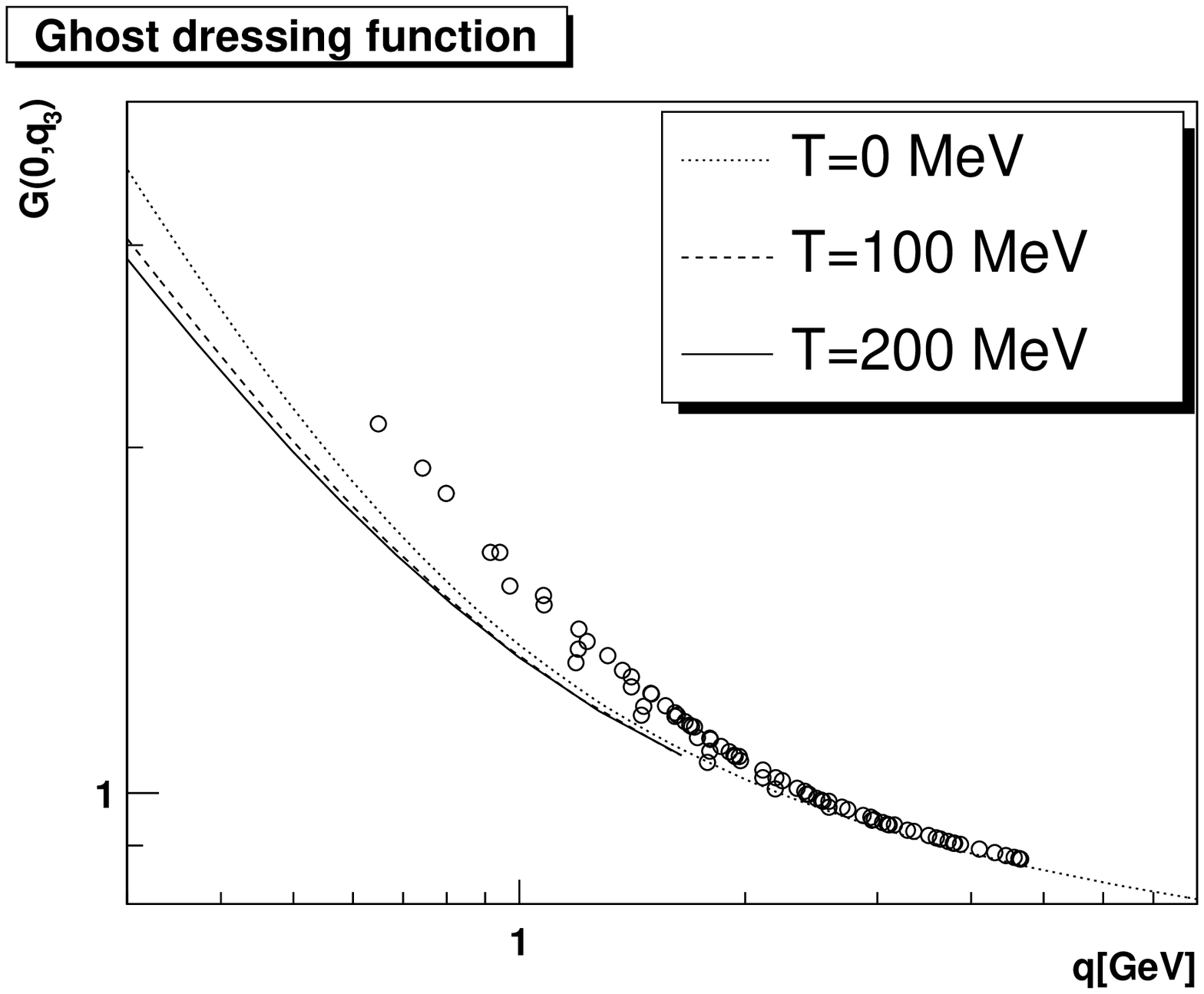}\epsfxsize=2.3in\epsfbox{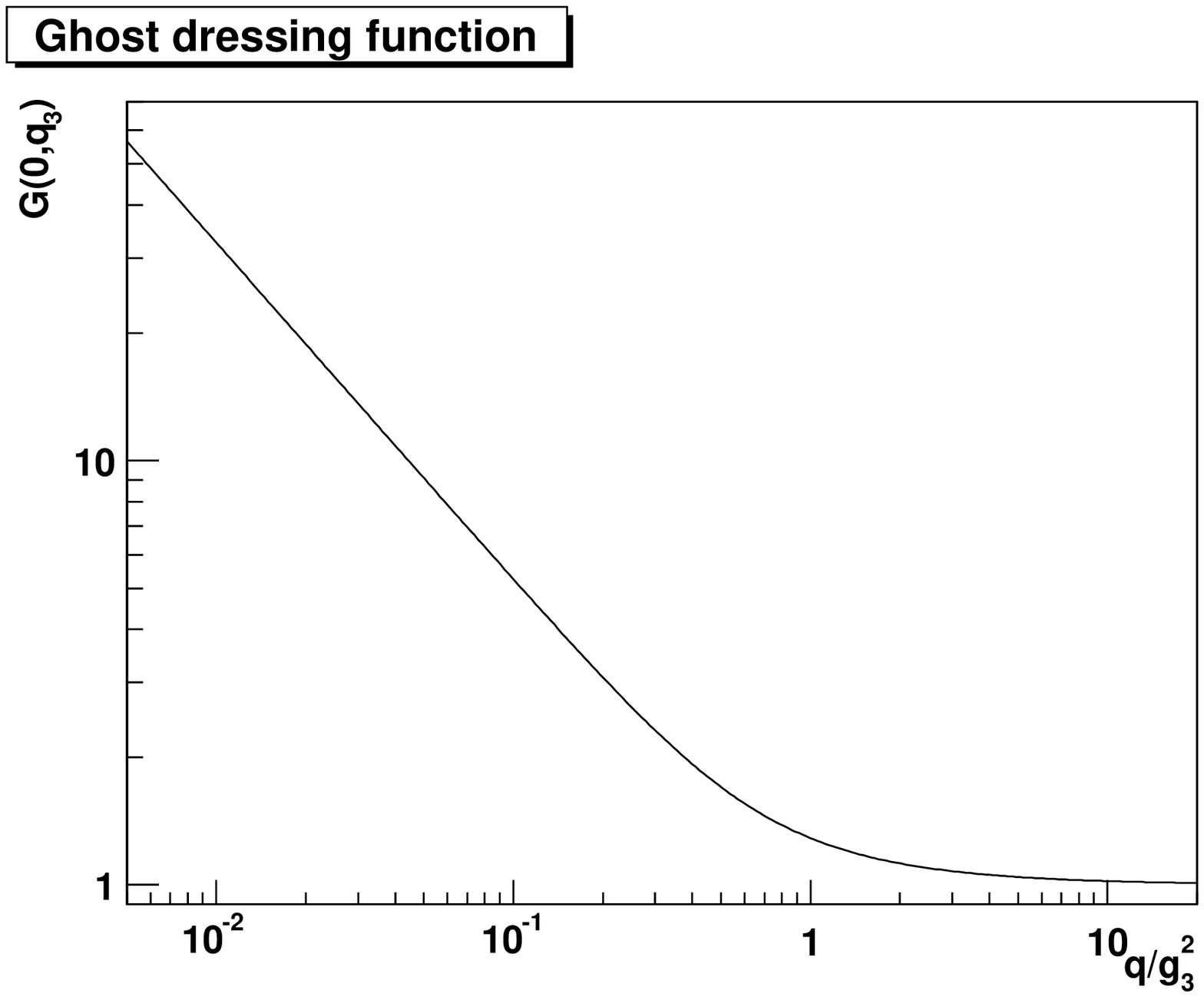}}   
\caption{The ghost dressing function compared to lattice 
results\protect\cite{Langfeld:2002dd}: 
The left panel is for $T<T_c$ while the right panel shows $T\to\infty$. $g_3$ is the 
dimensionful coupling constant in the dimensionally reduced theory.}
\label{ghostfig}
\end{figure}

Solutions for the ghost dressing function are shown in Fig~\ref{ghostfig}. It 
is found that they are not affected by temperature qualitatively. The divergence at 
zero momentum indicates the mediation of long-range forces which are still present 
at $T\to\infty$. In addition, this divergence is connected to the confinement 
mechanism\cite{Alkofer:2000wg} and thus indicates the presence of some residual
gluon confinement 
in the high-temperature phase.

\begin{figure}[ht]
\centerline{\epsfxsize=2.3in\epsfbox{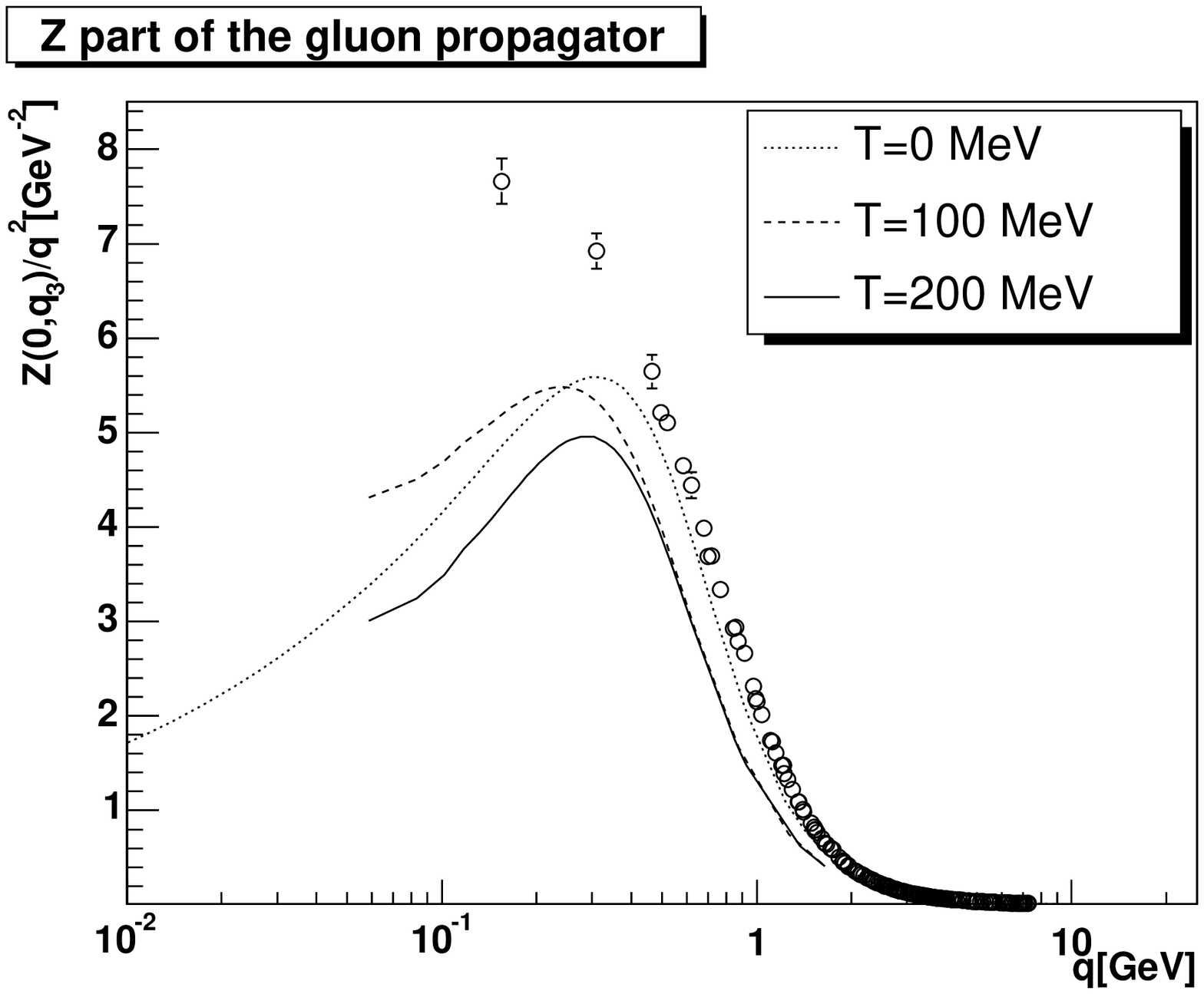}\epsfxsize=2.3in\epsfbox{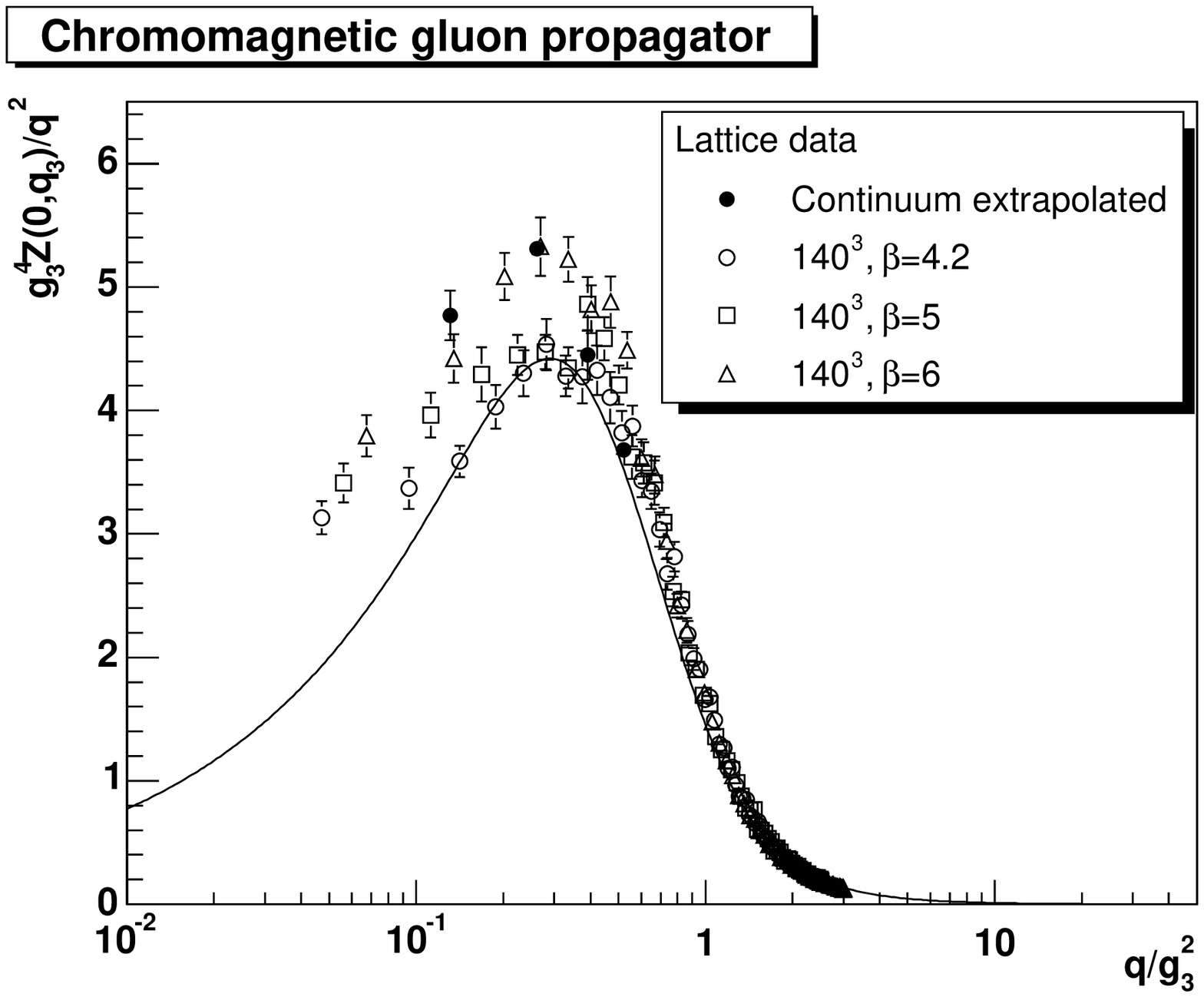}}   
\caption{The transverse gluon propagator: 
The left panel displays $T<T_c$ results compared 
to $T=0$ lattice data\protect\cite{Bowman:2004jm} 
while the right panel shows the $T\to\infty$ 
case compared to corresponding lattice data\protect\cite{Cucchieri:2001tw,Cucchieri:2003di}.
\label{gluonfig}}
\end{figure}

The results for the transverse gluon are shown in Fig.~\ref{gluonfig}. Below the phase 
transition, the propagator exhibits confinement by virtue of condition \eqref{confcrit}. 
It becomes steeper in the infrared with increasing temperature. In the high-temperature 
phase there is no qualitative difference. Hence the transverse gluon propagator also 
shows explicitly the presence of residual confinement which can be interpreted as 
over-screening. In fact, the temperature dependence is very slight. The lattice data 
at $T=0$ show strong finite-volume effects in the infrared and are expected to bend 
over for larger volumes as is the case in the dimensionally reduced theory, 
where significantly larger lattices can be employed.

\begin{figure}[ht]
\centerline{\epsfxsize=2.3in\epsfbox{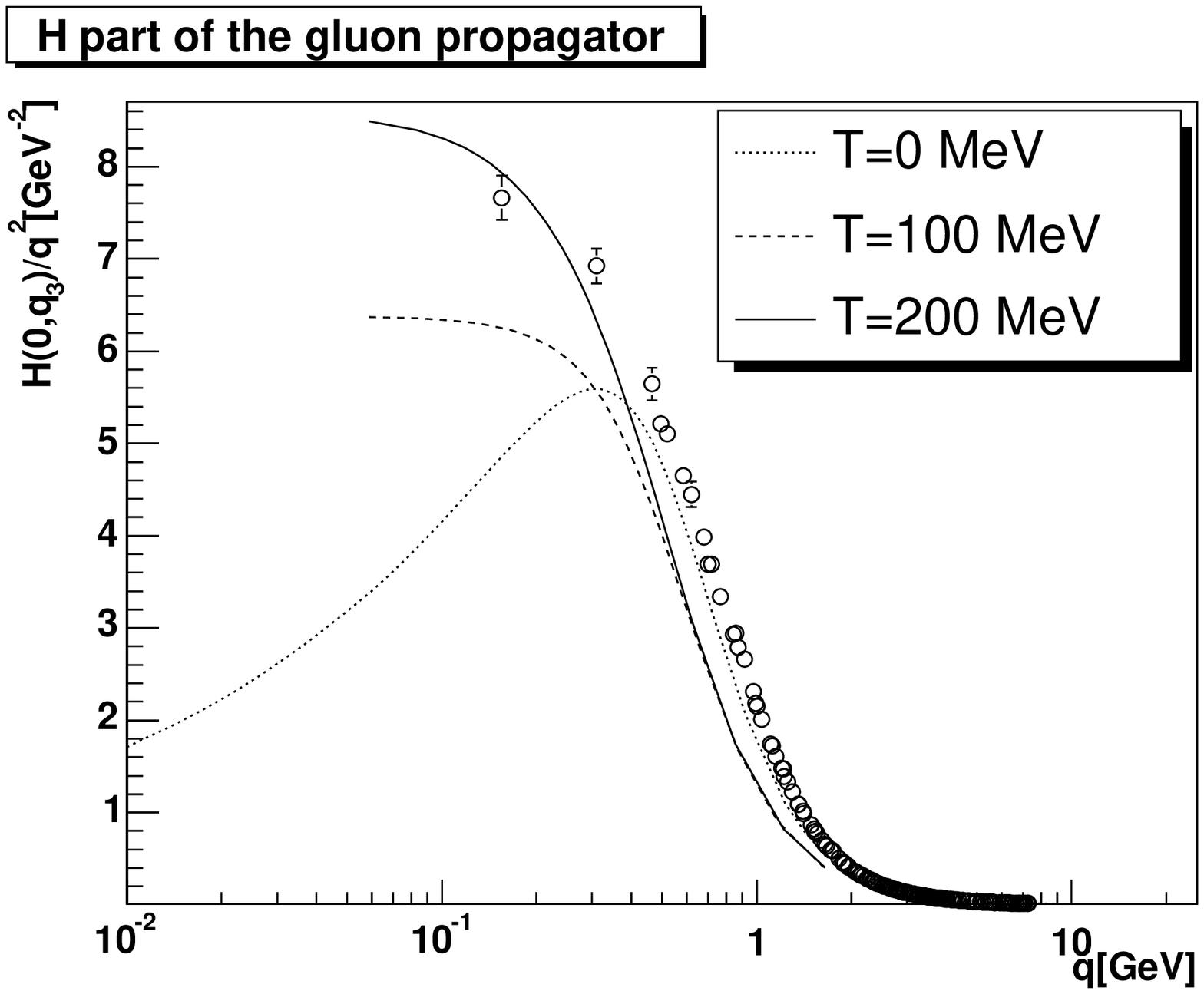}\epsfxsize=2.3in\epsfbox{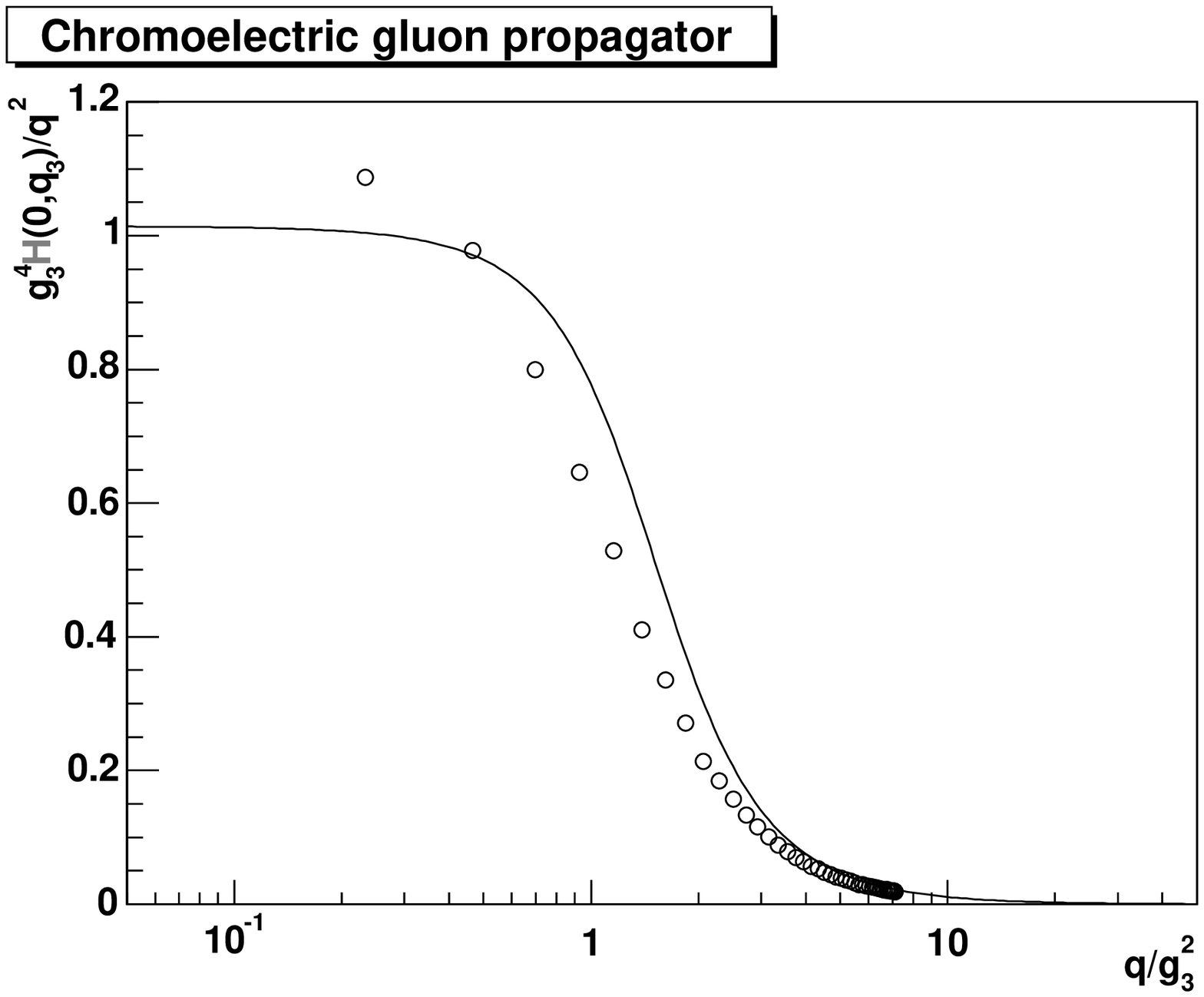}}   
\caption{The longitudinal gluon propagator: 
The left panel is for $T<T_c$ compared to $T=0$ 
lattice data\protect\cite{Bowman:2004jm} while in the right panel 
$T\to\infty$ results are compared to corresponding lattice 
data\protect\cite{Cucchieri:2001tw}. \label{higgsfig}}
\end{figure}

For the longitudinal gluon propagator the results are shown in Fig.~\ref{higgsfig}. 
At $T=0$, $H=Z$ and thus longitudianl gluons are confined as well. At non-zero temperature, 
$H$ becomes shallower in the infrared, and is thus severely affected by finite-volume 
effects. It, however, still exhibits confinement according to volume-scaling
studies\cite{torus}.
In the high-temperature phase, the situation is much different. In contrast to the 
transverse sector, it shows screening and behaves similar to a massive particle. 
Nevertheless a comparison to lattice results and perturbation theory 
indicates that the Higgs propagator contains sizeable non-perturbative 
effects\cite{d3}.

\section{Conclusions}

We have analyzed the infrared behavior of gluon and ghost propagators employing
Dyson-Schwinger equations.  At vanishing and small temperatures the results
show manifestly gluon confinement.  It is found that, even in the $T\to\infty$
limit, strong long-range correlations are present, leading to a
non-perturbative behavior of the soft modes. 
Part of the gluons are still confined: The Gribov-Zwanziger scenario (see {\it
e.g.\/} Zwanziger\cite{Zwanziger:2003cf}) applies at {\bf all} temperatures. 
These results, together with the one of lattice calculations, 
demonstrate the presence of strong non-perturbative effects even at high
temperatures. In the infrared sector of a Yang-Mills theory soft
quantum fluctuations win over thermal fluctuations at all temperatures.

\section*{Acknowledgments}
A.M. thanks the organizers of the Strong and Electroweak Matter 2004 for the very 
inspiring conference and the opportunity to present this work.

\end{document}